**Title:**

Coherent electron-spin-resonance manipulation of three individual spins in a triple quantum dot

**Authors:**

A. Noiri[1,2], J. Yoneda[1,2], T. Nakajima[1,2], T. Otsuka[1,2], M. R. Delbecq[1,2], K. Takeda[1,2], S. Amaha[2], G. Allison[2], A. Ludwig[3], A. D. Wieck[3], and S. Tarucha[1,2]

**Affiliations:**

*[1]Department of Applied Physics, University of Tokyo, 7-3-1 Hongo, Bunkyo-ku, Tokyo 113-8656, Japan*

*[2]RIKEN, Center for Emergent Matter Science (CEMS), Wako-shi, Saitama 351-0198, Japan*

*[3]Lehrstuhl für Angewandte Festkörperphysik, Ruhr-Universität Bochum, D-44780 Bochum, Germany*

**Abstract:**

Quantum dot arrays provide a promising platform for quantum information processing. For universal quantum simulation and computation, one central issue is to demonstrate the exhaustive controllability of quantum states. Here, we report the addressable manipulation of three single electron spins in a triple quantum dot using a technique combining electron-spin-resonance and a micro-magnet. The micro-magnet makes the local Zeeman field difference between neighboring spins much larger than the nuclear field fluctuation, which ensures the addressable driving of electron-spin-resonance by shifting the resonance condition for each spin. We observe distinct coherent Rabi oscillations for three spins in a semiconductor triple quantum dot with up to 25 MHz spin rotation frequencies. This individual manipulation over three spins enables us to arbitrarily change the magnetic spin quantum number of the three spin system, and thus to operate a triple-dot device as a three-qubit system in combination with the existing technique of exchange operations among three spins.

Semiconductor quantum dots (QDs) are appealing systems to implement solid-state qubits using confined electron spins [1-4]. Key technologies for operating the spin qubits have been realized to date using double QDs (DQDs), such as spin-state initialization and readout by the Pauli exclusion effect [1,2], single-electron spin resonance (ESR) [1,5-10] and exchange control [2,8,10]. They have been used to implement two spin-1/2 qubits [8,10] and a single Singlet-Triplet qubit [2]. More recently beyond DQDs triple and quadruple QDs (TQDs and QQDs) have been fabricated to implement an exchange-only qubit [3,11,12] and coupling between Singlet-Triplet qubits [13]. The electronic states are well controlled for these QDs [14-17], but full control of more than two spin-qubits, of any type, has been elusive although the QD architecture is predicted to have potential scalability. Demonstration of three qubits is an inevitable step not only for scaling up the qubit system but also for performing practical quantum algorithms, including quantum error-correcting codes [18,19] and quantum teleportation [20]. This will be achieved by the individual manipulation over three spins in combination with the existing technique of exchange operations [3,11,12].

In this work, we demonstrate coherent manipulation of three individual spin-qubits in a TQD with a proximal micro-magnet (MM). In the conventional ESR experiments in DQDs [1,5-9] the initialization and readout of the spins relies on Pauli-spin-blockade (PSB) under a finite source-drain bias. However, this scheme is hard to apply to initialize spins in linearly coupled TQDs because PSB is observed only in limited conditions [21]. Therefore, we instead initialize the spin using the so-called slow adiabatic passage [2,3,11,12]. We detect three ESR spectra corresponding to individual spins whose splittings (~ a few hundred MHz) are large enough for addressable manipulation [22]. We observe distinct coherent Rabi oscillations in a TQD with up to 25 MHz spin rotation frequencies. These results establish the guidelines for applying the MM-based ESR technique to systems with more than two electron spins.

Our device is a gate-defined lateral TQD fabricated in a two-dimensional electron gas (2DEG) formed at a GaAs/AlGaAs heterointerface 100 nm below the surface (Fig. 1(a)). We deposit a 250 nm Cobalt MM on top of the TQD gate electrodes with a 50 nm thick insulator in between. The MM is designed to create a local magnetic field with two important parameters for addressable three-spin rotations [9,22,23]. One is a difference $\Delta B_Z$ in the local magnetic field parallel to the external magnetic field $B_{ext}$, between neighboring QDs, which is ~50 mT in our case. This splits the ESR spectra and enables individual manipulation of three single spins. The other is a slanting field (~1 T/μm in our case) which mediates rapid rotation of spins [9]. All measurements are performed at a base temperature around 30mK and with an in-plane $B_{ext}$ ranging from 0.5 T to 1 T applied along the [110] crystalline axis (see Fig. 1(a)).

Fig. 1(b) shows the relevant TQD stability diagram in the few-electron regime, where the ($N_L$, $N_C$, $N_R$)

= (1,1,1) charge state is neighbored by the (2,0,1) and (1,0,2) states. Here $N_{L(C, R)}$ indicates the number of electrons inside the left (center, right) QD. We apply voltage pulses to gates PL and PR to rapidly control the energy-level detuning $\varepsilon$ between the singly-occupied state (1,1,1) and the doubly-occupied states, (2,0,1) and (1,0,2). Unlike previous experiments on exchange-interaction-based spin manipulation in TQDs [3,11,12,24], we tune the TQD such that the boundaries between (1,1,1) and the two doubly-occupied states are sufficiently separatead in energy. Then, the effect of the exchange couplings between neighboring QDs is dominated by the large (∼ 50 mT) $\Delta B_Z$ near the center of the (1,1,1) region where we manipulate spins by ESR. Three spin eigenstates are then defined by the Zeeman field in the same way as three isolated spins. On the other hand, near the boundary with one of the doubly-occupied states, we can approximate the TQD as a composite system of a DQD consisting of the left (right) and center QDs and an almost isolated right (left) QD. This is because in this regime the exchange coupling between the center QD and either the left or right QD spin overwhelms the Zeeman energy difference, while the interaction with the other QD spin is negligible due to the large energy detuning. In this way we can initialize and read out TQD spins with a detuning pulse in a manner analogous to that used with DQDs (Fig. 1(c)) [2,25]. For simplicity, in the following descriptions of TQD operations, we will use notations for DQD spin states such as $|S\rangle$, and $|T_{\pm,0}\rangle$ for singlet and triplet states, respectively, in the neighboring two QDs. We do not initialize and read out the third spin in this work because it is decoupled and does not affect our results. Nevertheless, the initialization and readout of the decoupled third spin should be possible under a sufficiently large external magnetic field [15, 26, 27].

Rapid (∼µs) preparation in the ground doubly-occupied singlet state $|S\rangle$ is achieved by moving near the (1,0,1) charge boundary (I in Figs. 1(d) and (e)), where the electron is exchanged with one in the lead. During the detuning ramp to the (1,1,1) region for state initialization (from M to O in Fig. 1(c)), the ground state $|S\rangle$ undergoes two anticrossings (Fig. 1(c)). The first one is between the $|S\rangle$ and $|T_+\rangle$ states mediated by the local in-plane magnetic field difference between neighboring QDs [28,29] (we denote the gap size as $\Delta_{ST+}$). The second one is between singly- and doubly-occupied singlet states hybridized by the inter-dot tunnel coupling $t$. These anticrossings, with different gaps, enable us to tune the spin state we prepare and measure just by changing the detuning rate. If the detuning ramp is adiabatic at the first anticrossing, we can initialize spins in $|T_+\rangle$ (yellow dashed curve in Fig. 1(c)). This is realized with a slow detuning ramp (> 10 µs in our case). On the other hand, if we non-adiabatically cross the first anticrossing, but ramp adiabatically with respect to the second one and $g\mu_B\Delta B_Z$, we can initialize spins in $|\downarrow\rangle|\uparrow\rangle$ (black dashed arrow in Fig. 1(c)), which is an eigenstate determined by local Zeeman energies. This so-called slow adiabatic passage [2,30] is achieved with 0.01 to 1 µs ramps in our case, given the relation among these energy scales: $\Delta_{ST+} \ll g\mu_B\Delta B_Z \ll t$ [25]. This condition is readily achieved by making $t$ large enough in our device because the MM is designed to make $g\mu_B\Delta B_Z$ much larger than $\Delta_{ST+}$. We use $|\downarrow\rangle|\uparrow\rangle$ initialization to conduct the ESR experiment because high fidelity of state initialization is more readily obtained compared to that of $|T_+\rangle$.

To demonstrate the feasibility of PSB-based spin state readout for ESR signal detection, we evaluate the life time of $|T_+\rangle$. It can be checked by continuously measuring the population decay after preparing $|T_+\rangle$ with a slow detuning ramp and returning to the measurement point M by non-adiabatically crossing the $S$-$T_+$ anticrossing. Figs. 1(d) and (e) show the stability diagrams obtained with the pulse sequence running continuously. We find the blockade life time to be strongly dependent on device parameters [31], and after optimization it can be enhanced up to 233 μs. With rf-detected charge sensing, we can resolve two charge states of (1,1,1) and (2,0,1) ((1,0,2)) in (d) ((e)) in 10 μs, and this serves as a single-shot readout of the neighboring spin correlation of a three spin system [3,32].

To demonstrate ESR of three individual spins, we initialize the spin state to $|\downarrow\rangle|\uparrow\rangle$ with a 1 μs ramp detuning pulse and measure the probability of detecting $|S\rangle$, $P_S$, after the reverse pulse (Fig. 2(a)). The probability $P_S$ decreases only at the resonance condition. Fig. 2(b) and (c) shows the measured $P_S$ as a function of the $B_{ext}$ and the MW frequency $f_{MW}$ while we apply sufficiently long (1 μs) MW bursts. Two parallel lines appear in each field-frequency plot. Those lines are assigned to resonance signals from respective QDs, which demonstrate the ESR manipulation of three individual spins. As seen from Fig. 2(d), the resonance condition of the center QD spin is the same, up to the size of nuclear field fluctuations in GaAs-based QDs [35,36], even when we use different doubly-occupied states for readout and initialization (Fig. 2(b) and (c)), as long as the operation point is common.

The observed ESR splittings may also come from the g-factor difference [37] as well as from the MM stray field. To quantify the contributions, we evaluate the g-factors and the local Zeeman fields $B_Z$ in individual QDs (Table 1) from the ESR spectra (Figs. 2(b), (c)). $B_Z$ is obtained by extrapolating the resonance line to zero frequency. We find local Zeeman splittings are nominally (∼86%) produced by the $\Delta B_Z$ at $B_{ext}$=1 T. The obtained values of $\Delta B_Z$ mostly agree with, but slightly deviate from the simulated local magnetic field property. A likely explanation for the deviation is a combination of the MM misalignment and the dislocation of electrons. The result can be explained, for example, by assuming a shift of the MM by 30 nm along the [1$\bar{1}$0] direction with a triangular electron arrangement as indicated by white circles in Fig. 2(e). This electron arrangement is reasonable because we apply a much more negative voltage to the center plunger (PC) gate than to the PL and PR gates.

Table 1 Summary of the electron g-factor and $B_Z$ measured from the ESR spectra. The design values are simulated using a boundary integral method [38]. Here we assume 180 nm inter-dot distance determined by the surface gate design. Variations after ± result from a 50 nm MM misalignment in both lateral directions.

|  | Fig. 2(b) | | Fig. 2(c) | |
|---|---|---|---|---|
|  | Left QD | Center QD | Center QD | Right QD |
| $|g|$ | 0.351 | 0.351 | 0.351 | 0.356 |
| $B_Z$ (experiment) | 20 mT | 78 mT | 76 mT | 161 mT |
| $B_Z$ (design) | 45±30 mT | 115±20 mT | 115±20 mT | 150±15 mT |
| $\Delta B_Z$ (experiment) | 58 mT | | 85 mT | |
| $\Delta B_Z$ (design) | 56±10 mT | | 33±6 mT | |
| Slanting field [a)] | 0.64 T/μm | 0.80 T/μm | 0.80 T/μm | 1.26 T/μm |

a) Calculated for each QD configuration shown as white circles in Fig. 2(e).

Next, we control the MW burst time $t_{MW}$ to see Rabi oscillations of single electron spins in each QD [1,5-10]. We observe distinct Rabi oscillations of the three electron spins in the TQD, with the Rabi frequency up to $f_{Rabi}$ = 25 MHz as shown in Figs. 3(a)-(c). We find that our data are fitted well with a Gaussian envelope with the decay time $T_2^{Rabi}$, $A\exp(-(t_{MW}/T_2^{Rabi})^2)\cos(2\pi f_{Rabi}t_{MW}) + B$ [9,25], rather than a power-law envelope with an initial phase shift [39]. This is possibly because the sampling time (~1 s) for a single data point is comparable to the typical nuclear spin decorrelation time associated with the dephasing [25,40]. However, further investigations are necessary to make a conclusion. Figure 3(d) shows the MW amplitude dependence of $f_{Rabi}$ for each QD [8,9]. We obtain the fastest Rabi oscillation in the right QD where the slanting field is largest (see Table 1). Ideally, $f_{Rabi}$ should be in proportion to the slanting field [6,9,23], however, the obtained ratio of $f_{Rabi}$ (left:center:right~5:4:18) is not fully accounted for by the slanting field alone (left:center:right~4:5:8). This discrepancy may come from the asymmetric MW coupling to the motion of each QD. Indeed, the right spin is indicated to couple to the MW stronger than the other spins by observing smaller $T_2^{Rabi}$ for the right spin compared to the other spins under the same MW power [9]. The combination of the largest slanting field and stronger coupling to the MW may explain our observation that the right spin has by far the largest $f_{Rabi}$. The ESR splitting of our device (~300 or 400 MHz) is so large that addressable manipulation of three single spins would be feasible even with a 25 MHz Rabi frequency: $g\mu_B\Delta B_Z/h \gg f_{Rabi}$.

Lastly, we discuss the origin of the ESR driving field (effective $B_{ac}$) in our experiment. There are two possible mechanisms for coherent spin rotation in our setup, namely, a MM-induced stray magnetic field and the spin-orbit interaction (SOI). Our device is fabricated along the crystal orientation in which both contributions from Rashba and Dresselhaus terms of SOI are constructive to MM-mediated driving [5,9,41]. As a result, the effective $B_{ac}$ is in proportion to $b_{sl} + 2B_{ext}(l_\alpha^{-1} + l_\beta^{-1})$ [41]. Here $b_{sl}$ is a slanting field gradient (~T/μm) and $l_{\alpha(\beta)}$ is the spin-orbit length for the Rashba (Dresselhaus) term (~ tens of μm [5,6]). The contribution from the MM slanting field is then roughly 10 times stronger than that from the SOI under $B_{ext} \sim 1$ T where we perform the ESR experiment. Therefore, we conclude that our ESR is mainly mediated by the MM field.

In conclusion, we demonstrate coherent manipulation of three individual spins in a linearly coupled TQD with a MM. We tune the device to the three electron regime around the $(2,0,1) - (1,1,1) - (1,0,2)$ charge states and isolate each QD for addressable ESR driving while selectively coupling two of them for initialization and readout. We show that this TQD condition allows us to initialize and read out two-spin correlations of three spins in an analogous manner to DQDs and to observe ESR signals of three spins. We show that our spin manipulation is coherent through observation of Rabi oscillations for each spin with a maximum Rabi frequency up to 25 MHz. This technique is also applicable to systems with more than three electron spins, an important step toward implementing a larger spin qubit system.

**Acknowledgement:**

We thank the Microwave Research Group in Caltech for technical supports. Part of this work is financially supported by the ImPACT Program of Council for Science, Technology and Innovation (Cabinet Office, Government of Japan) and Project for Development Systems of the Ministry of Education, Culture, Sports, Science and Technology, Japan, and the IARPA project "Multi-Qubit Coherent Operations" through Copenhagen University. AN acknowledges support from Advanced Leading Graduate Course for Photon Science (ALPS). JY and TO acknowledge financial support from Incentive Research Projects. TN acknowledges financial support from JSPS KAKENHI Grant Number 25790006. TO acknowledges financial support from the Grant-in-Aid for Research Young Scientists B, Yazaki Memorial Foundation for Science and Technology Research Grant, Japan Prize Foundation Research Grant, Advanced Technology Institute Research Grant, the Murata Science Foundation Research Grant, Izumi Science and Technology Foundation Research Grant, CREST, JST. ST acknowledges financial support by JSPS, Grant-in-Aid for Scientific Research (No. 26220710). AL and ADW acknowledge gratefully support of Mercur Pr-2013-0001, DFG-TRR160, BMBF - Q.com-H 16KIS0109, and the DFH/UFA CDFA-05-06.

**Figures:**

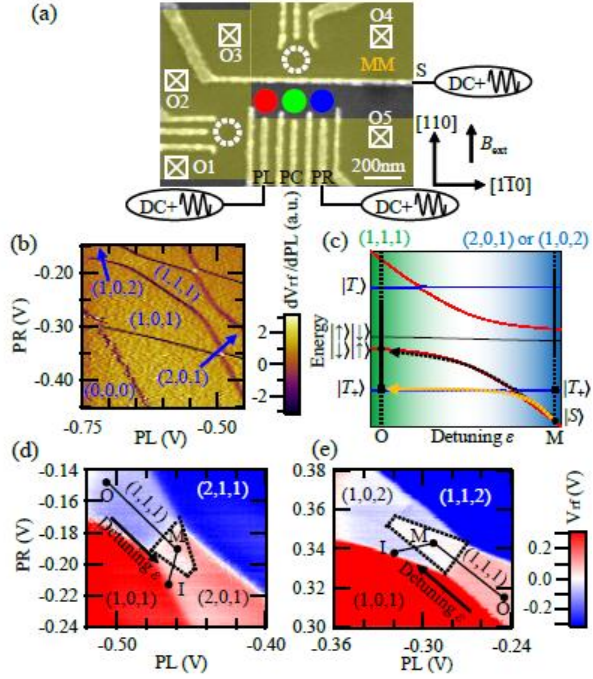

Fig. 1 (a) The scanning electron microscope image of a TQD device made in the same way as to that studied in this work. The QD positions are schematically illustrated by red, green and blue circles. The microwave (MW) to drive ESR is applied to gate S which is ideally equally coupled to all three QDs to suppress photon-assisted inter-dot tunneling [5]. Two plunger gates PL and PR are connected to an arbitrary waveform generator to control the QD conditions rapidly. The upper charge sensing QD is rf-detected ($V_{rf}$) at 206.5 MHz [33,34]. The MM design is shown in yellow. The wafer orientation and the direction of $B_{ext}$ are indicated by arrows. (b) Stability diagram of the few-electron regime measured by the upper rf-detected sensor conductance. (c) A simplified energy diagram of the three electron state in the DQD approximation as a function of detuning. Here $|S\rangle$ is a singlet-like spin state of a doubly-occupied left or right QD with a spin at the other end. Red curves are for singlet states hybridized due to the inter-dot tunnel coupling, while blue curves are for Zeeman-split triplet-like states (singly occupied). Around the center of the (1,1,1) region, the Zeeman splitting enhanced by the MM overwhelms the exchange coupling, so that the eigenstates are almost $|T_-\rangle$, $|\uparrow\rangle|\downarrow\rangle$, $|\downarrow\rangle|\uparrow\rangle$, $|T_+\rangle$. $|\downarrow\rangle|\uparrow\rangle$ is energetically lower than $|\uparrow\rangle|\downarrow\rangle$ because of the MM field. M (O) shows the measurement (operation) point. (d),(e) Pauli-blocked stability diagram around (1,1,1) and (2,0,1) in (d) ((1,0,2) in (e)) with the detuning pulse under $B_{ext}$=0.7 T. PSB is observed inside a trapezoidal region shown by black dashed lines with 30 μs time-averaged measurement just after returning to M.

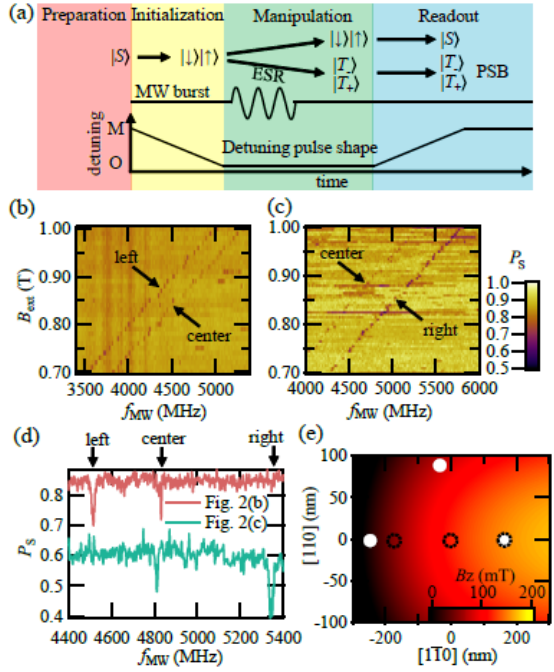

Fig. 2 (a) Pulse sequence used for ESR detection. A two spin state is initialized to $|\downarrow\rangle|\uparrow\rangle$. (b) ESR spectra operating around (1,1,1) and (2,0,1). (c) ESR spectra operating around (1,1,1) and (1,0,2). (d) ESR signal plotted in red (green) by tracing $P_S$ as a function of $f_{MW}$ at $B_{ext}$=0.9 T in (b) ((c)). The two data sets are offset by 0.2 for clarity. (e) Spatial distribution of the simulated $B_Z$. Black dashed circles show the designed positions of the three QDs. White circles show the QD positions assumed to explain the result well.

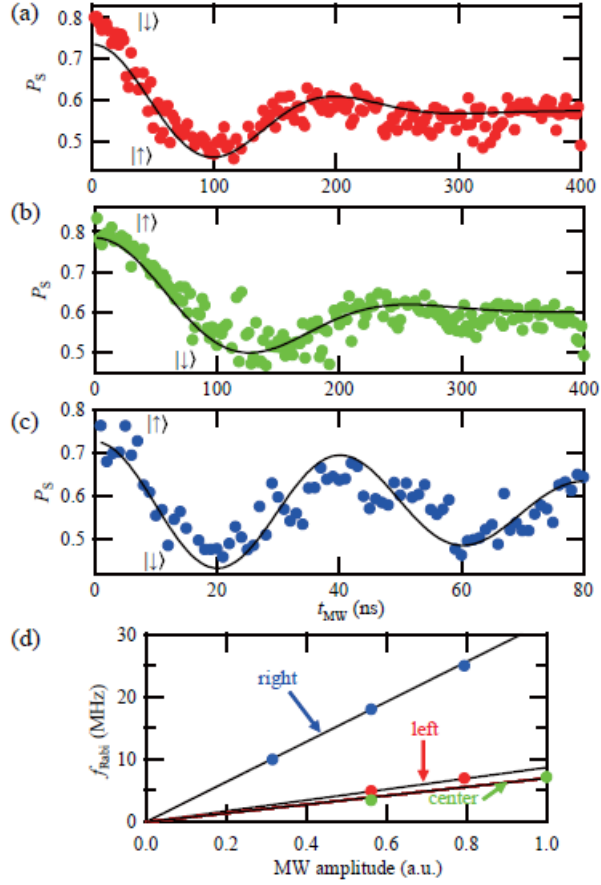

Fig. 3 Rabi oscillation of an electron spin in each QD and a fitting curve of $P_S = A\exp(-(t_{MW}/T_2^{Rabi})^2)\cos(2\pi f_{Rabi} t_{MW}) + B$ to the data points for (a) the left QD with $A$=0.16, $T_2^{Rabi}$=168 ns, $f_{Rabi}$=4.7 MHz, $B$=0.57, (b) center QD with $A$=0.18, $f_{Rabi}$=3.5 MHz, $T_2^{Rabi}$=176 ns, $B$=0.6, and (c) right QD with $A$=0.15, $f_{Rabi}$=24.6 MHz, $T_2^{Rabi}$=84 ns, $B$=0.58, respectively. In these measurements, we only change the MW burst length while keeping the detuning pulse parameters fixed. (a) and (b) are measured at the same condition with Fig. 2(b) where the spins in the left-center DQD are initialized to $|\downarrow\rangle|\uparrow\rangle$. On the other hand, (c) is measured at the same condition with Fig. 2(c). (d) MW amplitude dependence of $f_{Rabi}$. The electron spin in the right QD yields the largest $f_{Rabi}$.

Supplementary material:

Coherent electron-spin-resonance manipulation of three individual spins in a triple quantum dot

A. Noiri[1], J. Yoneda[1,2], T. Nakajima[1,2], T. Otsuka[1,2], M. R. Delbecq[1,2], K. Takeda[1,2], S. Amaha[2], G. Allison[2], A. Ludwig[3], A. D. Wieck[3], and S. Tarucha[1,2]

## 1. The validity of the DQD approximation

Here we discuss the required condition to validate our DQD approximation of the three-spin system in a collinear TQD at the initialization and readout step. Fig. S1(a) shows the three-spin energy diagram as a function of detuning [1-3] with symmetric inter-dot tunnel couplings $t$ between the neighboring QDs. Fig. S1(b) shows the zoom-in plot of Fig. S1(a). The only relevant change made by the full treatment is anti-crossing structures between doublet states indicated by black circles (see also Fig. 1(c) in the main text). An adiabatic detuning ramp over these additional anti-crossings will cause a state leakage to the other doublet, which is not included in the DQD approximation. The justification of our approximation is therefore the non-adiabaticity of the pulse at these anti-crossings. As seen from Fig. S1(c), the gap energy between doublet states ($\Delta_{D'D}$) becomes rapidly small with reducing $t$. Therefore, the condition for a slow adiabatic passage, $\Delta_{ST+}, \Delta_{D'D} \ll g\mu_B\Delta B_Z \ll t$, is readily achieved by tuning $t$ in an appropriate range (~GHz in our experiment). In this case, spins in two neighboring QDs out of the three will be initialized (and read out by the reverse pulse) as predicted by the DQD approximation as indicated by black curves in Fig. S1(b).

## 2. Fitting the Rabi oscillations

Next we discuss the detailed scheme for the fitting of the Rabi oscillations. We measure $P_S$ as a function of $f_{MW}$ and $t_{MW}$ to obtain the Rabi oscillations in a similar manner to those described in the supplementary material of ref 4. Here we sweep $f_{MW}$ (50 MHz around the resonance frequency) and then step $t_{MW}$ to take the 2D plot with a single data point acquisition time of ~1 s, collecting the minimum values of $P_S$ from each sweep. This scheme compensates for the nuclear field fluctuation between $f_{MW}$ sweeps with different values of $t_{MW}$. Therefore, the standard deviation of the Overhauser field $\sigma$ associated with the dephasing of the Rabi oscillations is determined by the sampling time of a single data point, which is estimated to be $\sigma$ ~1 MHz for ~1 s acquisition time [5]. Given the Rabi frequency of 3 MHz, our Rabi oscillations satisfy $f_{Rabi}/\sigma \gtrsim 3$, in which case the phase shift associated by the nuclear field fluctuation disappears [6] (see also Figs. S2(a)). To confirm this effect, we calculate $P_S$ as a function of $f_{MW}$ (in the range of $\pm 5f_{Rabi}$) and $t_{MW}$ (Chevron pattern) to extract the oscillation trace by taking a minimum value for each $t_{MW}$ with different values of $f_{Rabi}/\sigma$ as shown in Fig. S2(a). We find $\pi/4$ phase shift in the limit of $f_{Rabi}/\sigma \ll 1$, but as $f_{Rabi}/\sigma$ increases it disappears and the oscillation becomes saw tooth like [4]. This explains the Rabi oscillations of the left and center spins very well as shown in Figs. S2(b) and (c). Here we also take into account a relaxation process with $T_1 \sim 170$ ns, which is presumably due to the heating and photon-assisted tunneling under a strong MW power [6].

On the other hand, the 25 MHz Rabi oscillation in the right spin is fitted better by a sinusoidal function with a Gaussian envelope. This is because the range of the MW sweep $\Delta f_{\mathrm{MW}}$ is rather small compared to the Rabi frequency ($\Delta f_{\mathrm{MW}}/f_{\mathrm{Rabi}} \sim 1$). To simulate this situation, we calculate minima of $P_S$ only from the MW range of $\pm f_{\mathrm{Rabi}}$ around the resonance frequency, which fits the obtained Rabi oscillation well by taking into account a relaxation process with $T_1 = 84$ ns as shown in Fig. 2S(d). Given a sufficiently large Rabi frequency ($f_{\mathrm{Rabi}}/\sigma \gg 1$), the oscillation is approximated by the Gaussian decay function $P_S = A\exp(-(t_{\mathrm{MW}}/T_2^{\mathrm{Rabi}})^2)\cos(2\pi f_{\mathrm{Rabi}} t_{\mathrm{MW}}) + B$. Since this competition between the sampling time and the nuclear spin fluctuation is not the main scope of this paper, we use the sinusoidal function with the Gaussian envelope for fitting all the three Rabi oscillations presented in the main text to avoid confusion.

Figures for supplementary material

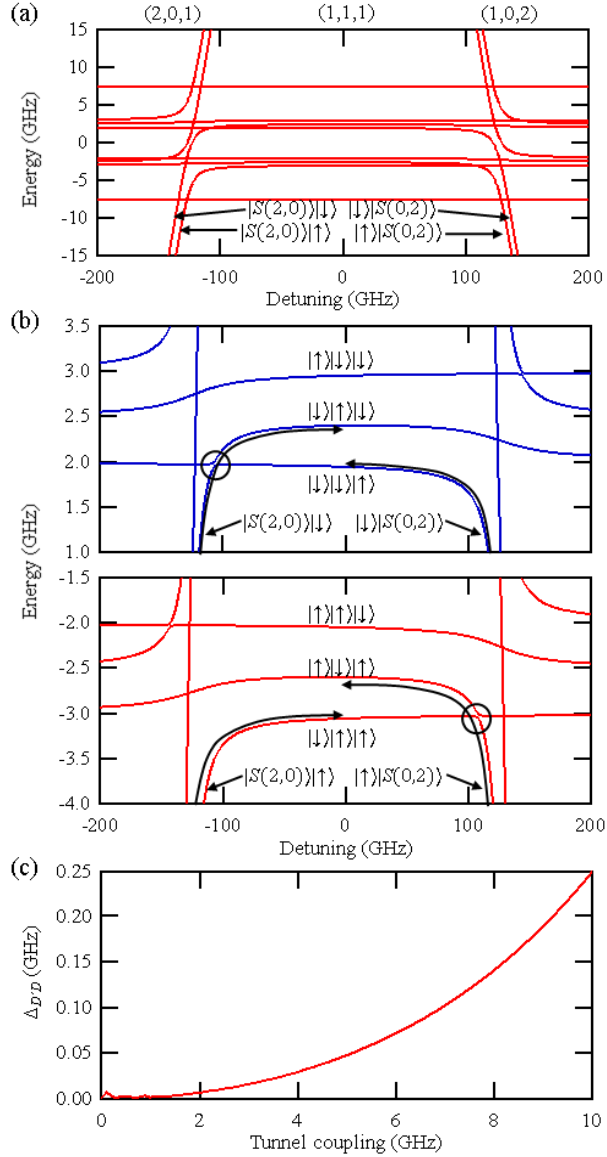

Fig. S1 (a) Energy diagram of the three spin states as a function of detuning $\varepsilon$. We use $\varepsilon_0 = \pm 125$ GHz (the size of the (1,1,1) region bounded by (2,0,1) and (1,0,2)), $t = 5$ GHz, $g\mu_B B_{ext} = 5$ GHz, $g\mu_B \Delta B_Z = 0.5$ GHz for the calculation parameters. (b) Zoom-in plot of (a) focusing on $S_z = -1/2$ (upper panel) and $S_z = 1/2$ (lower panel) states. The two-spin state of three spins is initialized to $|\downarrow\rangle|\uparrow\rangle|\sigma_z\rangle$ or $|\sigma_z\rangle|\downarrow\rangle|\uparrow\rangle$ states (with $\sigma_z = \uparrow$ or $\downarrow$) by a detuning ramp as indicated by black arrows. Our DQD approximation will be valid by non-adiabatically crossing the anti-crossing point between doublet states as indicated by the black circles. (c) $t$ dependence of $\Delta_{D'D}$ as shown by the black circles in (b).

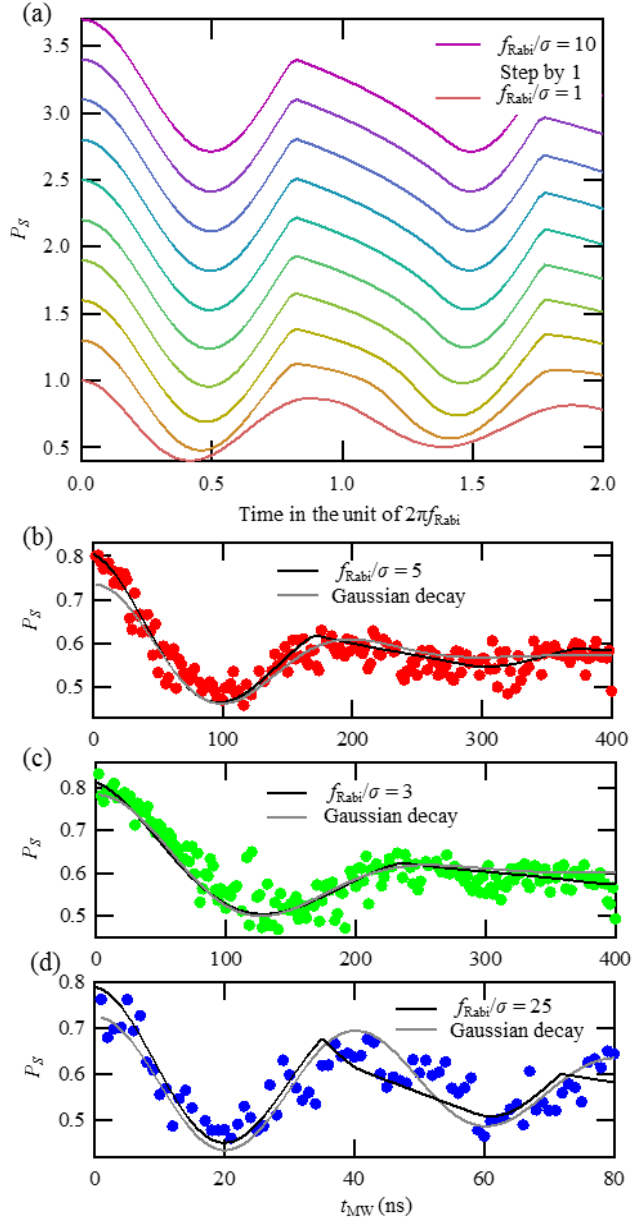

Fig. S2 (a) Calculated minima of $P_S$ extracted from Chevron patterns with different values of $f_{Rabi}/\sigma$. Each trace is offset by 0.3 for clarity. The phase shift associated by the nuclear field fluctuation disappears in $f_{Rabi}/\sigma \gtrsim 3$. (b)-(d) Rabi oscillations fitted by a saw tooth like trace. The fitting curve is generated from the calculated trace of (b) the left spin with $f_{Rabi}/\sigma$=5, (c) the center spin with $f_{Rabi}/\sigma$=3, (d) the right spin with $f_{Rabi}/\sigma$=25 by taking into account an exponential decay with (b) $T_1$=168 ns, (c), $T_1$=176 ns, (d) $T_1$=84 ns, respectively. We also show the fitting with the Gaussian decay function as shown in Fig. 3(a) for comparison.